\documentclass[prd,preprint,linenumbers,amssymb,amsmath,aps,showpacs,nofootinbib,superscriptaddress]{revtex4}
\usepackage{graphicx}
\usepackage{bm}
\usepackage{subfigure}
\usepackage{amssymb}
\usepackage{color,soul}
\usepackage{graphicx}
\graphicspath{{figure/}}
\DeclareGraphicsExtensions{.pdf,.png,.jpg}
\setulcolor{red}

\def\mdm{m_{_{\rm DM}}}
\def\sv{\langle \sigma v \rangle}

\newcommand{\beq}{\begin{equation}}
\newcommand{\eeq}{\end{equation}}
\newcommand{\beqa}{\begin{eqnarray}}
\newcommand{\eeqa}{\end{eqnarray}}

\begin{document}

\title{Constraints on box-shaped cosmic ray electron feature from
dark matter annihilation with the AMS-02 and DAMPE data}

\author{Lei Zu}
\affiliation{Key Laboratory of Dark Matter and Space Astronomy, Purple Mountain Observatory, Chinese Academy of Sciences, Nanjing 210008, China}
\affiliation{School of Astronomy and Space Science, University of Science and Technology of China, Hefei 230026, Anhui, China}
\author{Cun Zhang}
\affiliation{Key Laboratory of Dark Matter and Space Astronomy, Purple Mountain Observatory, Chinese Academy of Sciences, Nanjing 210008, China}
\affiliation{School of Physics, Nanjing University, Nanjing, 210092, China}
\author{Lei Feng\footnote{Corresponding author: fenglei@pmo.ac.cn}}
\affiliation{Key Laboratory of Dark Matter and Space Astronomy, Purple Mountain Observatory, Chinese Academy of Sciences, Nanjing 210008, China}
\author{Qiang Yuan\footnote{Corresponding author: yuanq@pmo.ac.cn}}
\affiliation{Key Laboratory of Dark Matter and Space Astronomy, Purple Mountain Observatory, Chinese Academy of Sciences, Nanjing 210008, China}
\affiliation{School of Astronomy and Space Science, University of Science and Technology of China, Hefei 230026, Anhui, China}
\author{Yi-Zhong Fan\footnote{Corresponding author: yzfan@pmo.ac.cn}}
\affiliation{Key Laboratory of Dark Matter and Space Astronomy, Purple Mountain Observatory, Chinese Academy of Sciences, Nanjing 210008, China}
\affiliation{School of Astronomy and Space Science, University of Science and Technology of China, Hefei 230026, Anhui, China}

\begin{abstract}
Precise measurements of spectra of cosmic ray electrons and positrons can
effectively probe the nature of dark matter (DM) particles. In a class
of models where DM particles initially annihilate into a pair of intermediate
particles which then decay into standard model particles, box-shaped
spectra can be generated. Such a kind of spectra are distinct from
astrophysical backgrounds, and can probably be regarded as characteristic
features of the DM annihilation. In this work, we search for such a feature
in the total electron plus positron spectrum measured by AMS-02 and DAMPE.
No significant evidence for such a DM annihilation component has been found.
The 95\% confidence level upper limits of the velocity-weighted annihilation
cross section are derived, which range from $\sim 10^{-26}~{\rm cm^3~s^{-1}}$
for DM mass of 50 GeV to $\sim 10^{-23}~{\rm cm^3~s^{-1}}$ for DM mass of
10 TeV.

\end{abstract}
\maketitle

\section{Introduction}

Dark Matter (DM) is one of the very important unresolved mysteries in
modern physics and astrophysics. The leading candidate of DM particle
is the so-called weakly interacting massive particle (WIMP).
Experiments with collider detection~\cite{DM_co,DM_co1}, direct
detection~\cite{DM_de,DM_de1} and indirect detection~\cite{DM_in,DM_in1}
have been carried out to search for WIMPs. The indirect detection method
aims to identify the relics in cosmic rays and/or gamma-rays from the
annihilation or decay of DM~\cite{box1,box2,box3,box4}.
Recent experiments such as PAMELA~\cite{pamela2} and AMS-02
\cite{ams02,ams02-2} discovered significant excess of cosmic ray positrons
on top of the conventional background model prediction, which stimulated
extensive discussion on the possible DM origin. At almost the same time,
excesses in the total electron plus positron spectrum were also reported
by several experiments~\cite{atic,fermi,ams02-3}, which may have a common
origin with the positron anomaly. The observations of the antiproton
fluxes are, however, largely consistent with the background model
prediction~\cite{pamela-antiproton1,pamela-antiproton2,ams-antiproton}
(see, however, \cite{cui2017,cuoco2017}).

The DM annihilation/decay models~\cite{Bringmann2008,Barger2009,Cirelli2009,
Yin2009,Zhang2009,yuan2015,Kopp2013,Cholis2013,jin2013} or new astrophysical
sources~\cite{yuksel,Hooper2009,Profumo2009,Malyshev2009,hu2009,Blasi,
Linden2013,Yin2013,diMauro2014,Fang2017} have been proposed to interpret
the electron and positron excesses. For DM annihilation models, the
annihilation cross section needs to be about $10^{-23}~{\rm cm^3 s^{-1}}$,
which is too large to account for the observed relic density unless a large
boost factor ($\sim 10^3$) is introduced~\cite{Cirelli2009,Nima2008}.
For decaying DM models, the lifetime of DM particles is about $10^{26}$~s.
However, both scenarios have already been stringently constrained by the
current multi-wavelength observationals~\cite{gammaconstrains1,
gammaconstrains2,gammaconstrains3,Bertone2009,Bergstrom2009,
cmbconstrains1,cmbconstrains2}.

With proper assumptions of the background contribution, the electron and
positron data can instead be used to constrain model parameters of DM,
in particular, for those models with very distinct spectral features
which are obviously missing in the data~\cite{bergstrom13}.
In this work, we focus on the specific scenario that DM particles
annihilate into intermediate particle pairs which then decay into standard
model particles~\cite{box6,box7}. Such a scenario is well motivated
and can provide a large boost factor \cite{Cirelli2009,Nima2008}.
It can also produce a box-shaped spectrum to get distinguished from
other astrophysical process effectively. This model has been adopted to
interpret the possible excess of the AMS-02 antiproton data at $\sim 400$
GeV \cite{box8}. The box-shaped $\gamma$-ray feature was also investigated
with the Fermi-LAT data \cite{box9,box10}.

Here we search for such spectral feature in the cosmic ray electron
(and/or positron) spectrum. In particular, we will employ the most
recent precise measurement of the total electron (i.e., electron plus
positron) spectrum by the DArk Matter Particle Explorer (DAMPE;
\cite{dampe-detector0,dampe-detector}). DAMPE is a high-energy cosmic
ray particle detector with unprecedentedly high energy resolution and an
excellent hadron rejection capability~\cite{dampe-detector0,dampe-detector}.
It can hence measure the total electron spectrum with high precision,
low background, and small systematic uncertainty. Recently, the DAMPE
collaboration published the first result about the measurement of the
electron plus positron spectrum from 25 GeV to 4.6 TeV~\cite{dampe-data}.
The wide energy coverage and high precision of the data can significantly
improve the constraints on the properties of DM particles~\cite{Yuan2017}.

This work is organized as follows: In Section II, we present the spectrum
of electrons generated by the two-step cascade annihilation of DM.
In Section III we briefly describe the propagation and background of
electrons and positrons. In Section IV A, we assume the annihilation takes
place in the whole Milky Way halo and perform the search for such signal
in the AMS-02 and DAMPE data.
In Section IV B we examine the scenario of the presence of a local
DM subhalo.
We summarize our results with some discussion in Section IV.

\section{Box-Shaped spectra}\label{sec::box}

We focus on the two-step annihilation model, in which DM particles
first annihilate into a pair of scalar particles $\phi$ which then decay
into electrons and positrons. In the rest frame of $\phi$, the energy of
the final electrons and positrons is $E_{e^\pm}=m_{\phi}/2$. Since we
focus on high-energy cosmic rays in this work, it is reasonable to assume
that $m_e\ll m_{\phi}<\mdm$, where $\mdm$ is the mass of DM particles.
If the mass of $\phi$ is comparable with $\mdm$ or $2m_e$, a narrow
peak spectrum of final state electrons and positrons would be produced
\cite{box8}. In the lab frame, DM particles are non-relativistic and
$E_{\phi}\approx\mdm$, hence the energy of electrons or positrons is
\begin{eqnarray}
E=\frac{m_{\phi}^2+\beta \cos \theta \sqrt{4\mdm^2 m_e^2 \left(\beta^2 \cos^2 \theta -1 \right) +m_{\phi}^4 }}{2\mdm\left( 1-\beta^2 \cos^2 \theta \right)},
\end{eqnarray}
where $\beta=\sqrt{1-m_\phi^2/\mdm^2}$, $\theta$ is the angle between
the outgoing electron/positron and the parent scalar in the lab frame.
For scalar intermediator $\phi$, the distribution of outgoing angles
of electrons/positrons is isotropic in the rest frame of $\phi$.
Therefore the electrons/positrons would have a box-shaped spectrum
in the lab frame as
\begin{eqnarray}
\frac{dN}{dE}=\frac{2}{\sqrt{\left(\mdm^2-m_{\phi}^2\right)\left(1-4m_e^2/m_\phi^2 \right)}}\,\Theta(E-E_-)\,\Theta(E_+-E),
\label{dnde}
\end{eqnarray}
where $\Theta$ is Heaviside step function and $E_{\pm}=\mdm/2\pm\sqrt
{\left(\mdm^2-m_{\phi}^2\right)\left(m_{\phi}^2-4m_e^2\right)}/2m_\phi$.
Our results reduce to that of Refs.~\cite{box6,box7} if we set
$dN/dE_{\rm rest}=\delta(E_{\rm rest}-{m_\phi}/{2})$ in the rest frame
of the intermediator. If we set $m_e=0$, Eq.~(\ref{dnde}) gives the photon
spectrum due to the two-step annihilation process~\cite{box9}. Note that
we have neglected the electroweak corrections on the electron spectrum
because its effect is tiny. In Ref.~\cite{PPPC4}, the authors calculated
the full spectrum with the electroweak corrections using the Pythia
package~\cite{pythia}, assuming a light MeV-scale intermediate particle.
We will compare our results with theirs (labelled as ``PPPC4'') in the
following.

\begin{figure}[htbp]
\centering
\includegraphics[width=0.6\columnwidth]{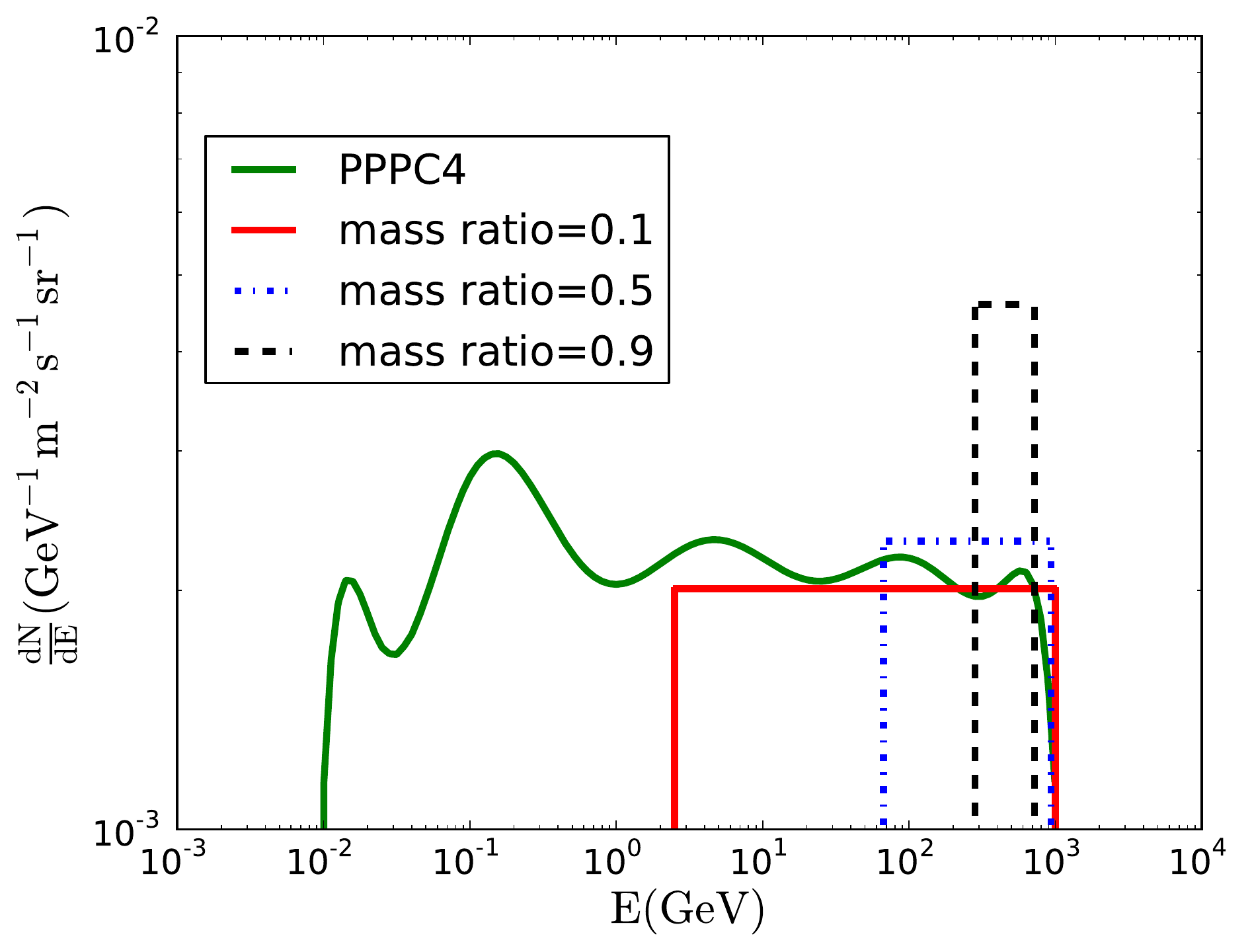}
\caption{Energy spectra of $e^+e^-$ at production for different mass ratios
between $m_{\phi}$ and $\mdm$. Also shown is that from PPPC4 (green line).}
\label{boxSED}
\end{figure}

There are two free parameters ($\mdm$ and $m_{\phi}$) determining the spectrum.
In the following we adopt $\mdm$ and the mass ratio $m_{\phi}/\mdm$ to present
the results. If $m_{\phi}/\mdm \sim 1$, the result reduces to a line spectrum.
On the other hand, if $m_{\phi}/m_{\rm DM}\ll 1$~(but still have $m_{\phi}
\gg m_e$), the produced box-shaped spectrum will be quite broad.
For illustration we show in Fig.~\ref{boxSED} the spectra of electrons for
$\mdm=1$ TeV and mass ratios of $(0.1, 0.5, 0.9)$, respectively. It is clear
to show that the spectrum is broader for smaller mass ratio. The result from
PPPC4 (with $\mdm=1$ TeV) is also presented in Fig.~\ref{boxSED}. The PPPC4 spectrum \cite{PPPC4} is a numerical result with a light mediator including the soft photon radiation process and the high order corrections. Its general shape is indeed box-like though it has wiggles.

\section{Propagation of electrons and the backgrounds}

\subsection{Propagation of cosmic ray electrons in the Galaxy}

Cosmic ray electrons and positrons propagate diffusively in the Galaxy,
with significant cooling due to synchrotron and inverse Compton radiation
processes. There are some numerical tools to compute the propagation of
cosmic rays, such as GALPROP~\cite{galprop} and DRAGON~\cite{dragon}.
Here we adopt the LikeDM package~\cite{likedm}, which is essentially
equivalent to GALPROP but employs a Green's function method (i.e., with
quasi-monochromatic injection spectrum at a series of energy grids) based
on numerical tables obtained with GALPROP for given spatial distribution
of the sources, to deal with the propagation of cosmic rays. This method
has been verified to give good approximation to the GALPROP output, but
is much more efficient~\cite{likedm}.

The propagation is assumed to work in a diffusion reacceleration framework,
with propagation parameters being determined by the Boron-to-Carbon ratio
data and the Fermi diffuse $\gamma$-ray emission~\cite{propagation}.
The main propagation parameters are: the diffusion coefficient $D(E)=\beta
D_0(E/4\,{\rm GeV})^{\delta}$ with $D_0=5.3\times10^{28}$~cm$^2$s$^{-1}$
and $\delta=0.33$, the half-height of the propagation cynlinder $z_h=4$ kpc,
the Alfvenic speed which characterizes the reacceleration effect $v_A=33.5$
km~s$^{-1}$. This set of propagation parameters is widely adopted in
literature as the canonical ``medium'' one. Since high energy electrons
and positrons can only travel a limited distance before cooled down,
we expect that choosing other propagation parameters with different $z_h$
values would not affect the results significantly.

\subsection{Backgrounds}

The astrophysical background includes the conventional primary electrons
from e.g., supernova remnants, and secondary electrons and positrons from
the inelastic collisions between cosmic ray nuclei and the interstellar
medium. Additional astrophysical sources, in particular pulsars, are also
expected to be high-energy electron/positron sources~\cite{pulsar}, and may
be responsible for the positron anomaly~\cite{yuksel,Hooper2009,Profumo2009}.
In this work we discuss two approaches of the backgrounds.

\subsubsection{Phenomenological background}

Since we are searching for spectral features which are distinct from
the ``smooth" background, it is reasonable to assume that the majority
of the observational data can be fitted by the background
\cite{bergstrom13,ibarra14}. Therefore a so-called ``minimum model'',
as that employed in Ref.~\cite{ams02}, is adopted. However, to better
reproduce the wide-band data from GeV to multi-TeV, we slightly extend
the ``minimum model'' with a low-energy break and a high-energy cutoff.
In this case the background model is directly fitted to the data,
without considering the propagation effect\footnote{For the DM component
to be discussed later, the propagation is still included.}.
The background model includes three components, the primary $e^-$,
secondary $e^+e^-$, and a source term of $e^+e^-$~\cite{likedm}, i.e.,
\begin{eqnarray}
\phi_{e^-}&=&C_{e^-}E^{-\gamma_1^{e^-}}\left[1+\left( E/E_{\rm br}^{e^-}\right)^{\gamma_2^{e^-}}\right]^{-1}{\rm exp}\left(-E/E^{e^-}_c\right),\\
\phi_{e^+}&=&C_{e^+}E^{-\gamma_1^{e^+}}\left[1+\left( E/E_{\rm br}^{e^+}\right)^{\gamma_2^{e^+}}\right]^{-1},\\
\phi_{\rm s}&=&C_{\rm s}E^{-\gamma^{s}}{\rm exp}\left(-E/E_{\rm c}^{\rm s}\right).
\end{eqnarray}

The total background energy spectrum of $e^-+e^+$ is then
\begin{equation}
\phi_{{\rm bkg},e^\pm}=\phi_{e^-} + 1.6\phi_{e^+} + 2\phi_{\rm s},
\end{equation}
where the factor 1.6 is due to the asymmetry of the electron and positron
productions in $pp$ collisions~\cite{Kamae2006}. The positron fraction is
\begin{equation}
f=(\phi_{e^+} + \phi_{\rm s})/\phi_{{\rm bkg},e^\pm}.
\end{equation}

The best-fit parameters are summarized in Table \ref{table:bkg}.
Note that they are somewhat different from that given in Ref.~\cite{likedm}
because of different data sets used in the fitting. When the DM contribution
is added in the model, we enable the backgrounds to vary to some degree
through multiplying adjustment factors
$\alpha_i E^{\beta_i}$, with $i=\{e^-,\,e^+,\,s\}$, on $\phi_e^-$,
$\phi_e^+$, and $\phi_{\rm s}$, respectively~\cite{Cirelli2009}.
Parameters $\alpha_i$ and $\beta_i$ are optimized during the fitting.

\begin{table}[!htb]
\caption {Best-fit parameters of the backgrounds}
\begin{tabular}{lccccc}
\hline \hline
 & $C$ & $\gamma_1$ & $\gamma_2$ & $E_{\rm br}$ & $E_c$ \\
 & (GeV$^{-1}$m$^{-2}$s$^{-1}$sr$^{-1}$) &  &  & (GeV) & (GeV) \\
\hline
$\phi_{e^-}$ & $21.3417$ & $0.8380$ & $2.4075$ & $3.3391$ & $1.4435\times10^4$ \\
$\phi_{e^+}$ & $1.1947$  & $0.7138$ & $2.5898$ & $2.7479$ & ... \\
$\phi_{s}$   & $0.9799$  & $2.3828$ & ... & ... & $842.93$ \\
\hline
\hline
\end{tabular}
\label{table:bkg}
\end{table}

\subsubsection{Physical background}

A more physical way to calculate the background starts with the injection
spectrum of different components of sources, and calculate the propagated
spectra of them. The LikeDM package is used to calculate the propagation
of various components. For the injection spectrum of primary electrons,
a three-segment broken power-law model with an exponential cutoff is assumed.
The first break at several GeV is to account for the low-energy data, and
the second break at several tens GeV is to explain the spectral hardening
\cite{feng2014,yuan2013,li2015}, and the cutoff is to reproduce the DAMPE
high-energy data \cite{dampe-data}. The injection spectrum of primary
electrons is then
\begin{eqnarray}
\Phi_{e^-}&=&A_{e^-}E^{-\nu_1^{e^-}}\left[1+\left( E/E_{\rm br1}^{e^-}
\right)^{3}\right]^{(\nu_1^{e^-}-\nu_2^{e^-})/3}
\left[1+\left( E/E_{\rm br2}^{e^-}\right)^{3}\right]
^{(\nu_2^{e^-}-\nu_3^{e^-})/3}\nonumber\\
&\times&{\rm exp}\left(-E/E_{c}^{e^-}\right).
\end{eqnarray}
Here we use $\Phi$ to describe the injection spectrum, in order to
distinguish from the propagated fluxes $\phi$ given in Eqs. (3-7).

The secondary positron spectrum from $pp$ collisions is calculated by
the GALPROP code. This component only needs to be calculated once.
A constant factor is multiplied to its flux during the fitting, which
accounts for possible uncertainties of the theoretical prediction
\cite{background}.

A pulsar-like component is also added in the model, which is expected
to contribute to the positron and electron excesses. The injection
spectrum of electrons and positrons from pulsars is described by an
exponential cutoff power-law form
\begin{eqnarray}
\Phi_{\rm psr}=A_{\rm psr}E^{-\nu^{\rm psr}}{\rm exp}\left(-E/E_c^{\rm psr}\right).
\end{eqnarray}
The spatial distribution of pulsars is adopted to be the same as the
primary cosmic ray source distribution~\cite{galprop}. Note that here
we assume a continuously distributed pulsar population to account for
the primary $e^+e^-$. It has been discussed extensively that nearby
isolated pulsars may be important to contribute to the
electron/positron fluxes around TeV energies~\cite{Profumo2009,Yin2013,
diMauro2014,Fang2017}. Given that there are many free uncertain parameters
in such a scenario, we simply assume a population of such sources in this
work. The spectra of electrons and positrons from a series of sources are
actually similar to that from a source population.


\section{Results}

We use a maximum likelihood fitting method based on the MINUIT tool in
ROOT to search for the DM component. The data used include the AMS-02
positron fraction~\cite{ams02}, the AMS-02 total electron fluxes below
25 GeV~\cite{ams02-3}, and the DAMPE data~\cite{dampe-data}. The background
parameters are optimized using the profile likelihood method. In this study we adopt the Navarro-Frenk-White (NFW) DM density profile~\cite{NFW}, which is
\begin{eqnarray}
 \rho _{NFW}(r)=\frac{\rho _s}{(\frac{r}{r_s})(1+\frac{r}{r_s})^2}.
\end{eqnarray}
The profile parameters can be found in Table II of Ref.~\cite{NFWparameters}.
\subsection{DM annihilation in the Milky Way halo}

\begin{figure}[htbp]
\centering
\subfigure[$m_{\phi}/\mdm=0.1$]{\includegraphics[width=0.45\columnwidth]{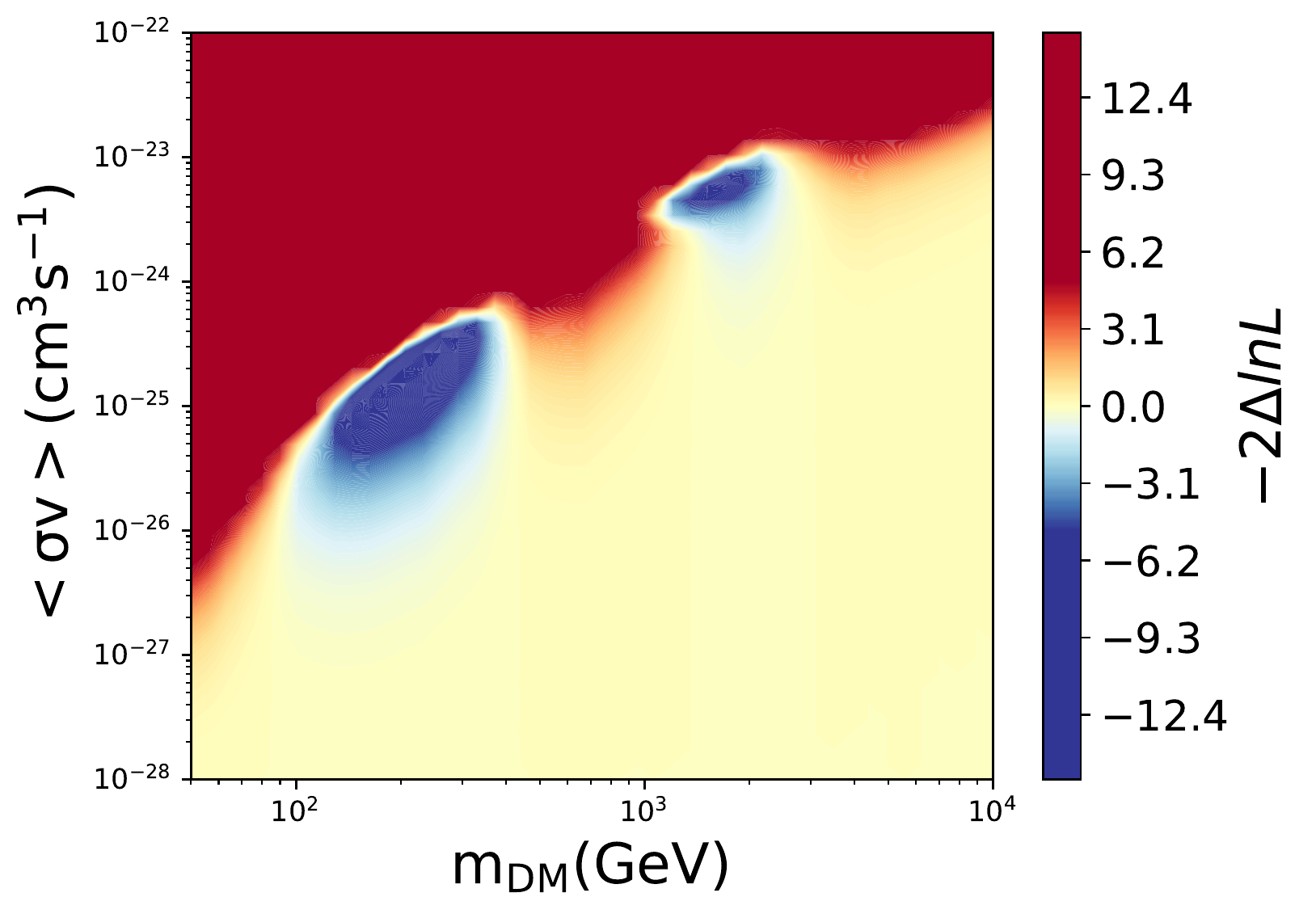}}
\subfigure[$m_{\phi}/\mdm=0.5$]{\includegraphics[width=0.45\columnwidth]{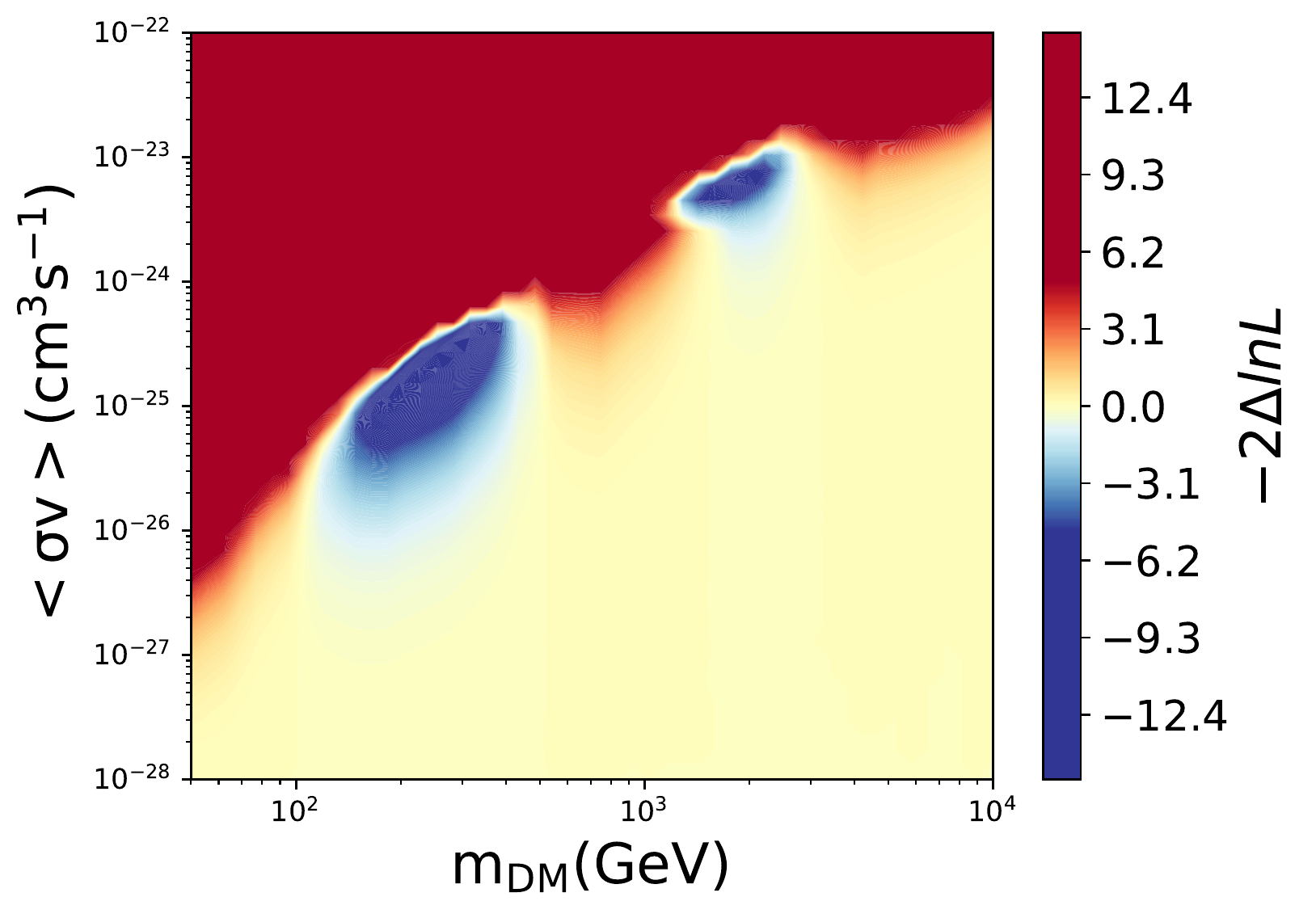}}
\subfigure[$m_{\phi}/\mdm=0.9$]{\includegraphics[width=0.45\columnwidth]{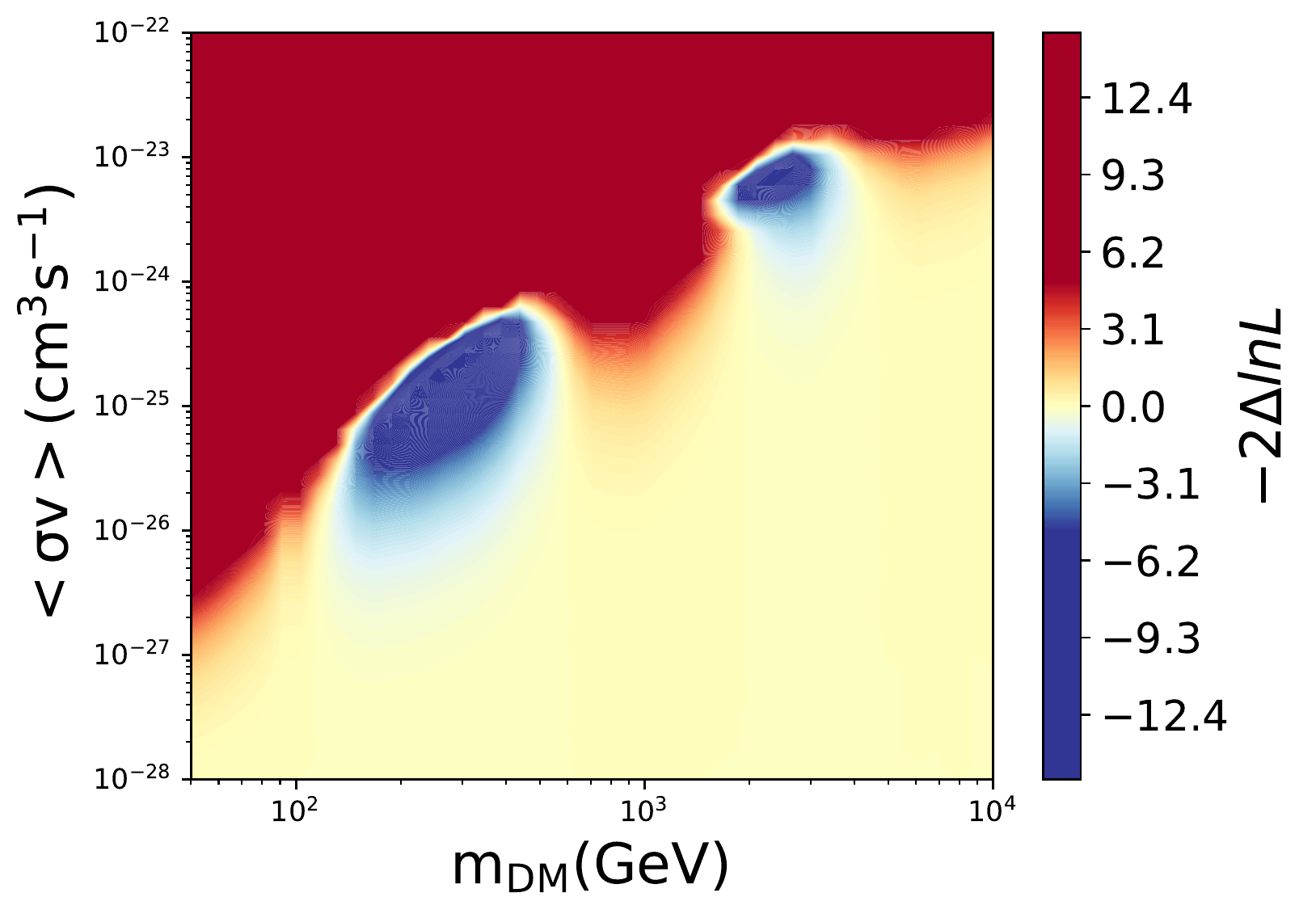}}
\subfigure[PPPC4]{\includegraphics[width=0.45\columnwidth]{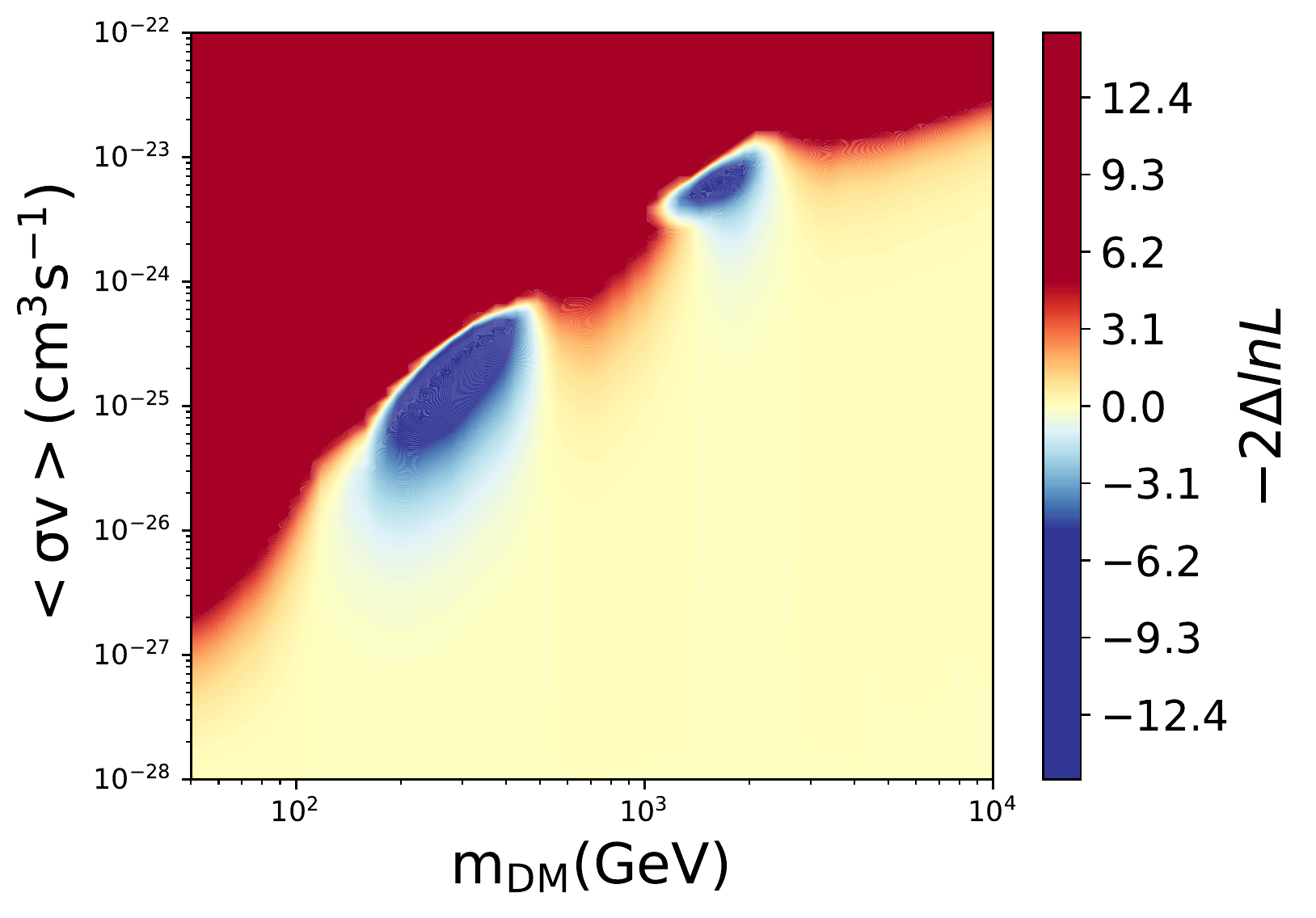}}
\caption{Log-likelihood maps on the $\mdm-\sv$ plane, for injection spectra
with different mass ratios ($m_{\phi}/\mdm=0.1,0.5,0.9$) and the PPPC4
spectrum. The physical background is assumed.}
\label{lmap}
\end{figure}

\begin{figure}[htbp]
\centering
\includegraphics[width=0.45\columnwidth]{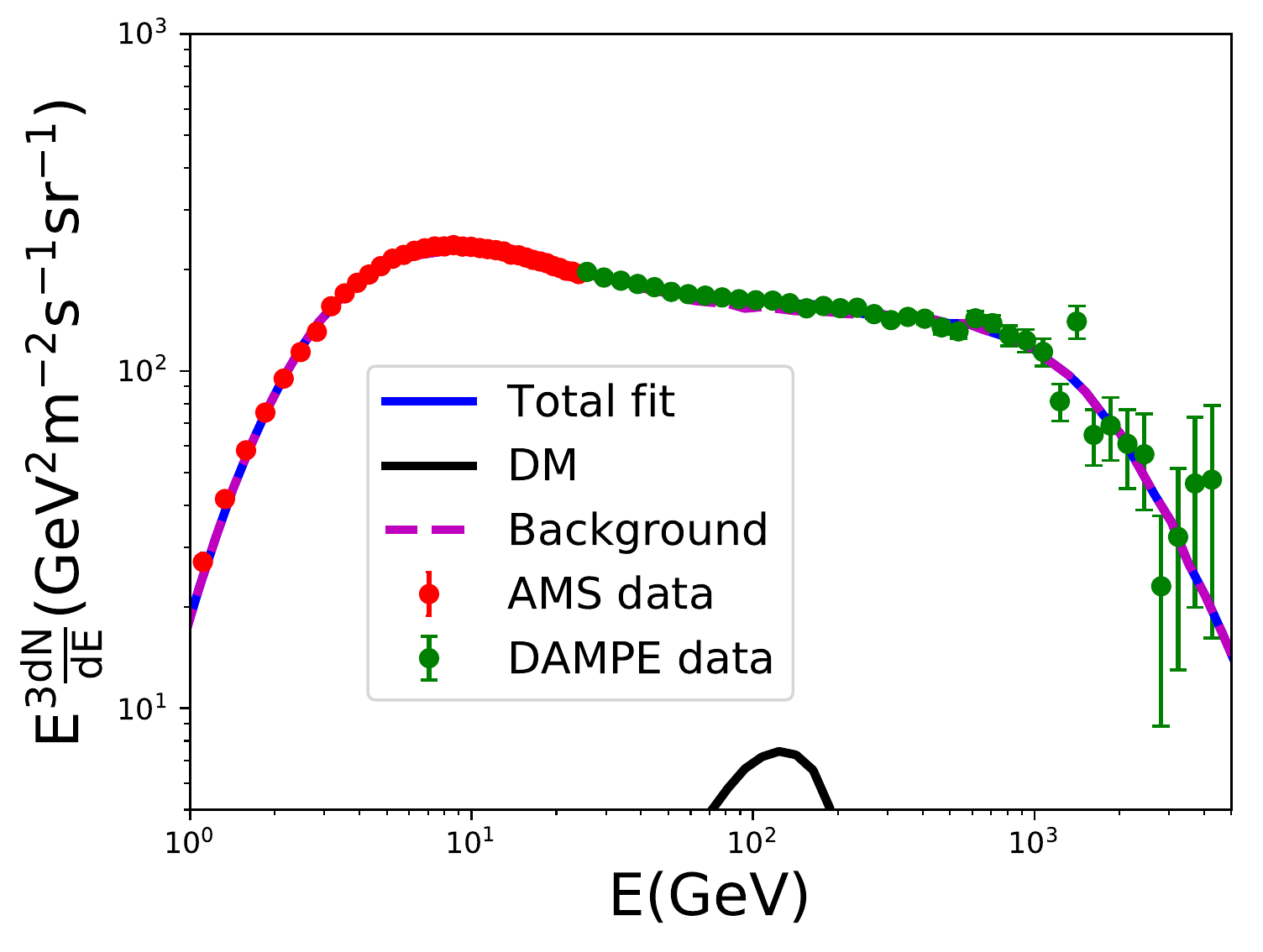}
\includegraphics[width=0.45\columnwidth]{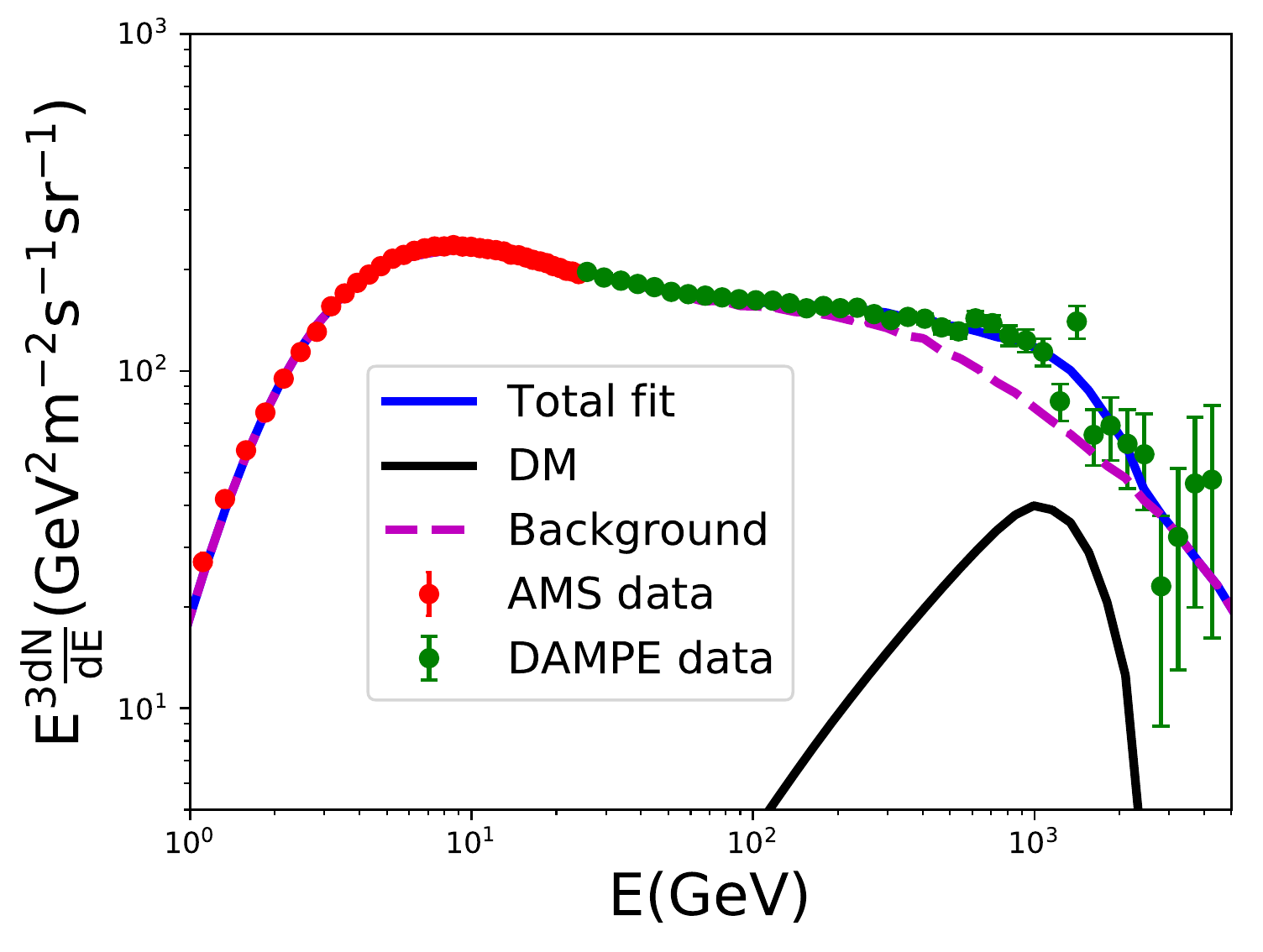}
\caption{Best-fit model prediction of the energy spectrum of electrons plus
positrons, for $\mdm=331$ GeV, $\sv=2.0\times10^{-25}~{\rm cm^3~s^{-1}}$
(left panel), and $\mdm=2870$ GeV and $\sv=7.9\times10^{-24}~{\rm cm^3~s^{-1}}$
(right panel). In both cases, $m_\phi/\mdm=0.9$ and the physical background
are adopted. Also shown are the data of AMS-02 \cite{ams02-3} and
DAMPE~\cite{dampe-data}.}
\label{fit}
\end{figure}

We first consider the scenario that DM annihilates in the whole Milky
Way halo. The maps of $-2\Delta\ln{\mathcal L}=-2(\ln \mathcal{L}
-\ln \mathcal{L}_0)$, where ${\mathcal L}_0$ (${\mathcal L}$) is the
likelihood without (with) the DM contribution, on the $(\mdm,\sv)$
plane for $e^+e^-$ injection spectra for different mass ratios
$m_{\phi}/\mdm$ and the PPPC4 energy spectrum are shown in Fig.~\ref{lmap}.
Here the physical background model is adopted. We find that there are
two regions (blue in Fig.~\ref{lmap}) with $\mdm\sim(100-500)$ GeV and
$\sim(2-3)$ TeV are ``favored'' by the data. Both regions are due to the
degeneracy between the DM contribution and the background (in particular,
the pulsar contribution). To better understand this, we show in
Fig.~\ref{fit} the comparison of the total electron spectrum between the
best-fit model and the data, for mass ratio of $m_{\phi}/\mdm=0.9$ and
$(\mdm,\sv)=(331~{\rm GeV},\,2.0\times10^{-25}~{\rm cm^3~s^{-1}})$
(left panel) and $(2870~{\rm GeV},\, 7.9\times10^{-24}~{\rm cm^3~s^{-1}})$
(right panel). The low mass region is favored when the pulsar component
has a very hard spectrum, which gives a ``dip'' on the spectrum and the
addition of a DM component can somehow improve the fit. We have tested
that if the pulsar injection spectrum is set to be softer than $E^{-1}$,
this low mass region becomes much less significant. The same applies to
the high mass region.

We also test the results for the phenomenological background model.
We find that only the $\sim(2-3)$ TeV region is slightly favored,
with $-2\Delta\ln{\mathcal L}$ about $-2$. This is mainly because the
phenomenological background model fits the data better than the
physical model, and the left room for DM contribution is smaller.
The $(2-3)$ TeV region again reflects the degeneracy between the DM
component and the $\phi_s$ component.

\begin{figure}[htbp]
\centering
\includegraphics[width=0.45\columnwidth]{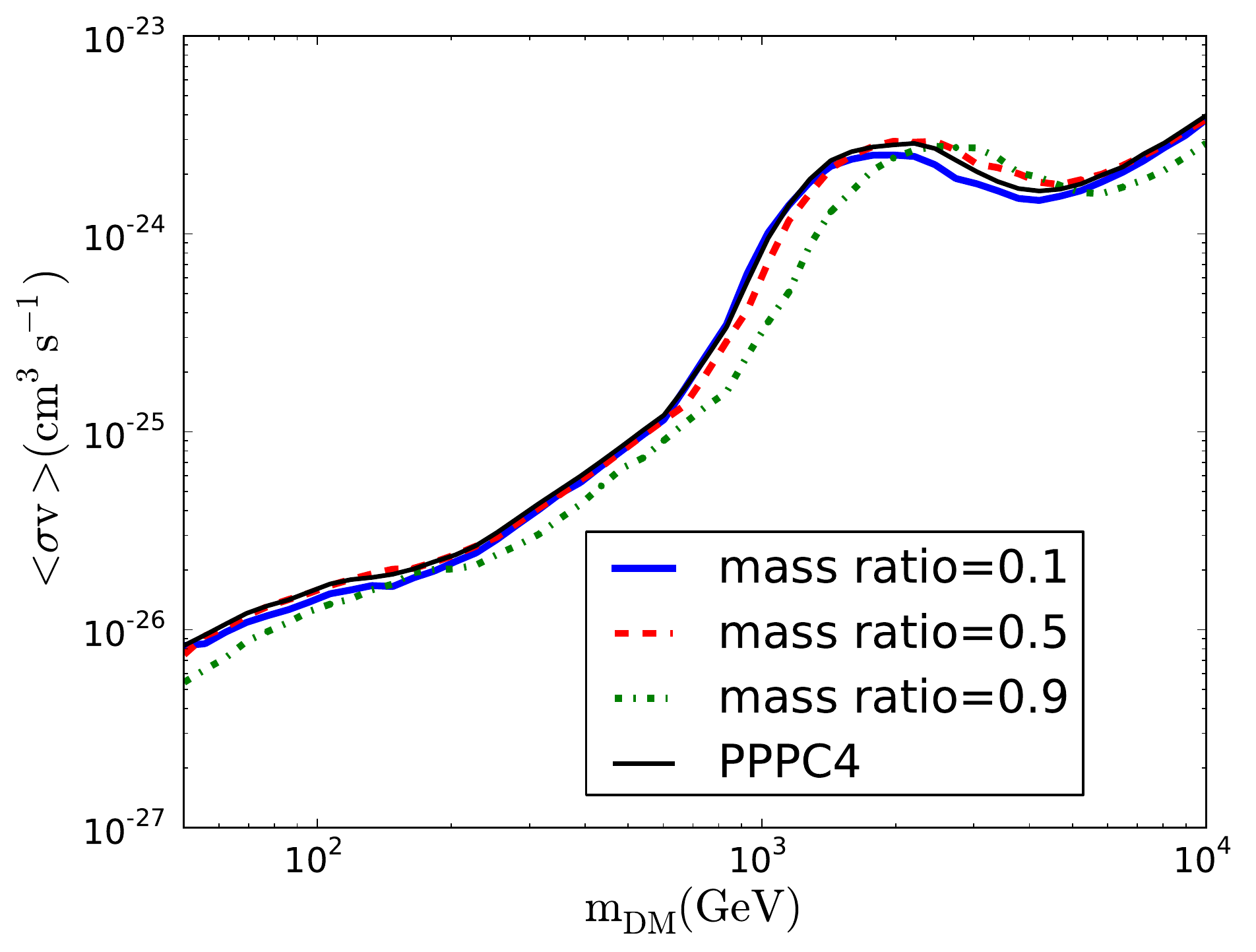}
\includegraphics[width=0.45\columnwidth]{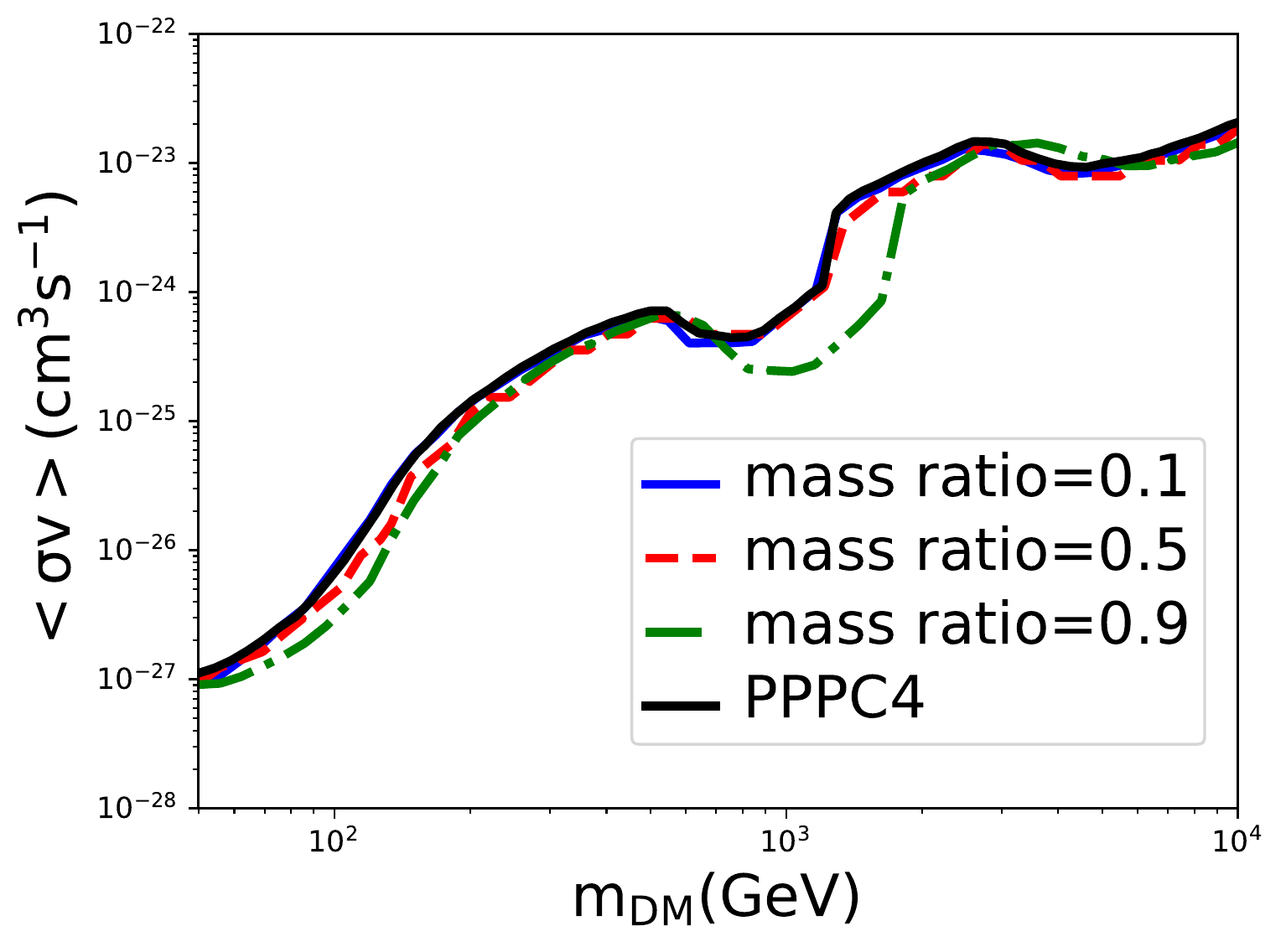}
\caption{The $95\%$ confidence level upper limits on the DM annihilation
cross section $\sv$ as a function of $\mdm$ for the phenomenological (left)
and physical (right) background models.}
\label{limit}
\end{figure}

\begin{figure}[htbp]
\centering
\includegraphics[width=0.8\columnwidth]{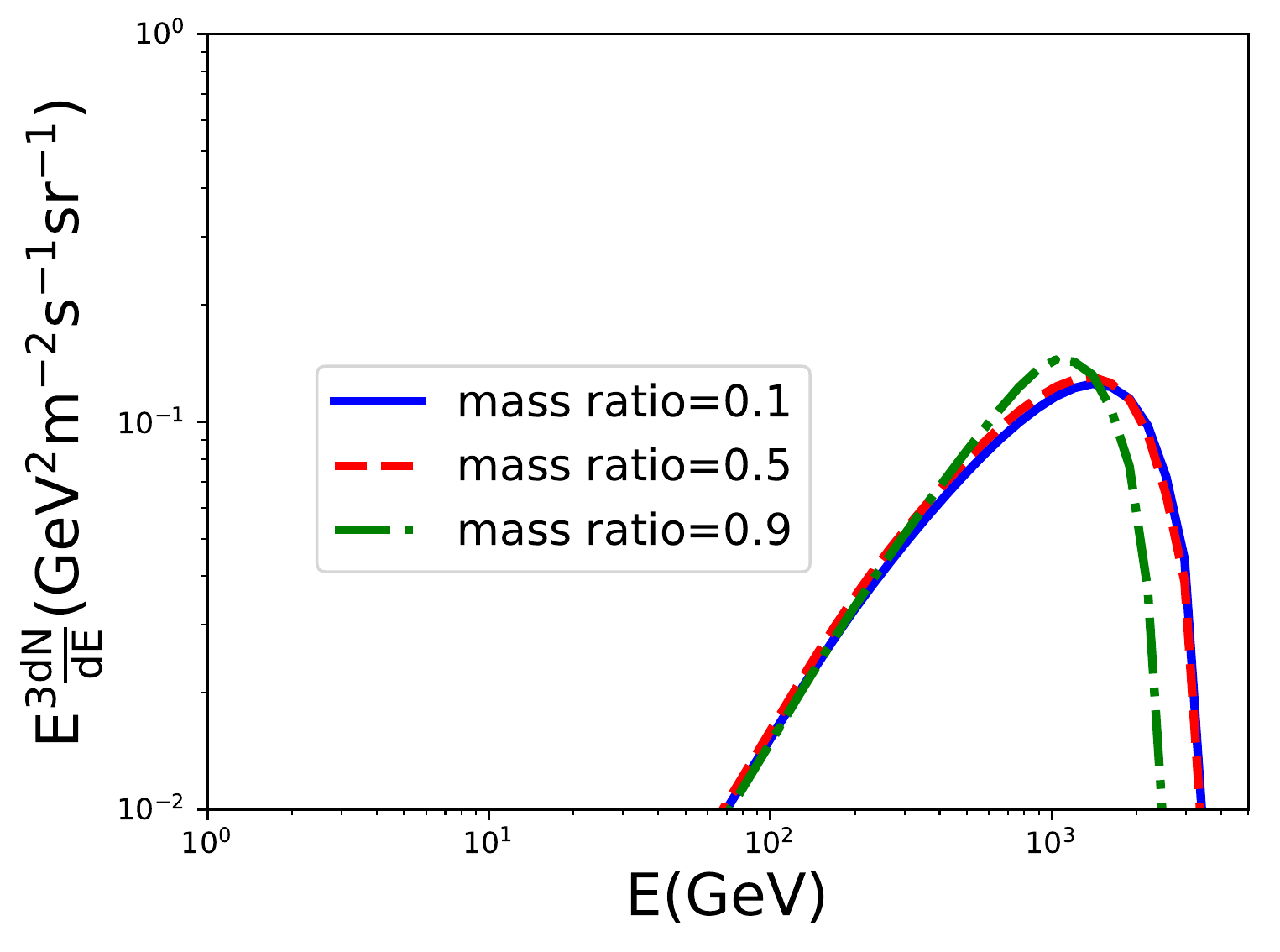}
\caption{Propagated spectra of electrons for $\mdm=3000$~GeV and
$m_{\phi}/\mdm=0.1,0.5,0.9$.}
\label{propagation}
\end{figure}

We thus give the 95\% upper limits of the annihilation cross section for
given DM mass $\mdm$, through setting $-\ln{\mathcal L}(\sv)\leq
-\ln{\mathcal L}_{\rm max}+1.35$. The results are given in Fig.~\ref{limit}.
The left panel is for the phenomenological background, and the right panel
is for the physical background. Since the results depend on the background
assumptions, we conservatively adopt the larger values as the final
constraints. The upper limits range from $\sim 10^{-26}~{\rm cm^3~s^{-1}}$
for $\mdm=50~{\rm GeV}$ to $\sim 10^{-23}~{\rm cm^3~s^{-1}}$ for
$\mdm=10~{\rm TeV}$. The results for different mass ratios do not
differ significantly from each other. This is because the injection
spectra have been broadened after the propagation, and their differences
become smaller (see Fig.~\ref{propagation}).
For $m_\phi/\mdm=0.9$ the constraints are slightly tighter than the
others. This is expected since a higher mass ratio of $m_{\phi}/\mdm$
gives a narrower injection $e^+e^-$ spectrum (Fig.~\ref{boxSED}).
The constraints become weaker for $\mdm \sim 2$ TeV and $\sim 300$ GeV,
due to the ``favored signal'' regions in such mass ranges.

\subsection{DM annihilation in a local subhalo}

The DM annihilation into $e^+e^-$ channel in a local DM subhalo was
proposed to explain the tentative peak structure at 1.4 TeV of the DAMPE
data~\cite{Yuan2017,Huang2017,Jin2017}. If the mass ratio $m_{\phi}/\mdm$
is very close to 1, the resulting $e^+e^-$ spectrum from the DM annihilation
will be nearly monnchromatic. In this subsection we examine whether such
a narrow box-shaped $e^+e^-$ spectrum can reasonably reproduce the data.

\begin{figure}[htbp]
\subfigure[$m_{\phi}/\mdm=0.5$]{\includegraphics[width=0.45\columnwidth]{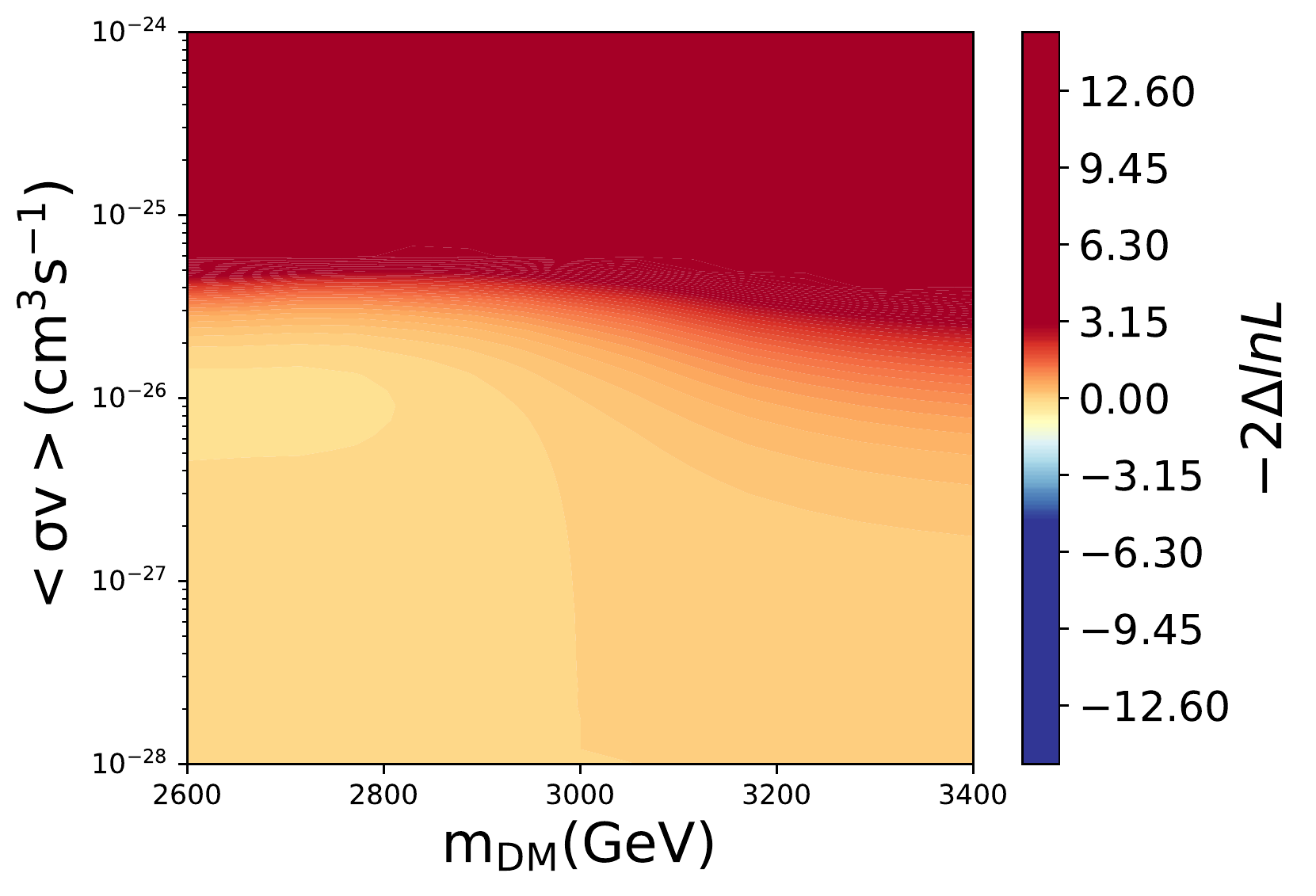}}
\subfigure[$m_{\phi}/\mdm=0.995$]{\includegraphics[width=0.45\columnwidth]{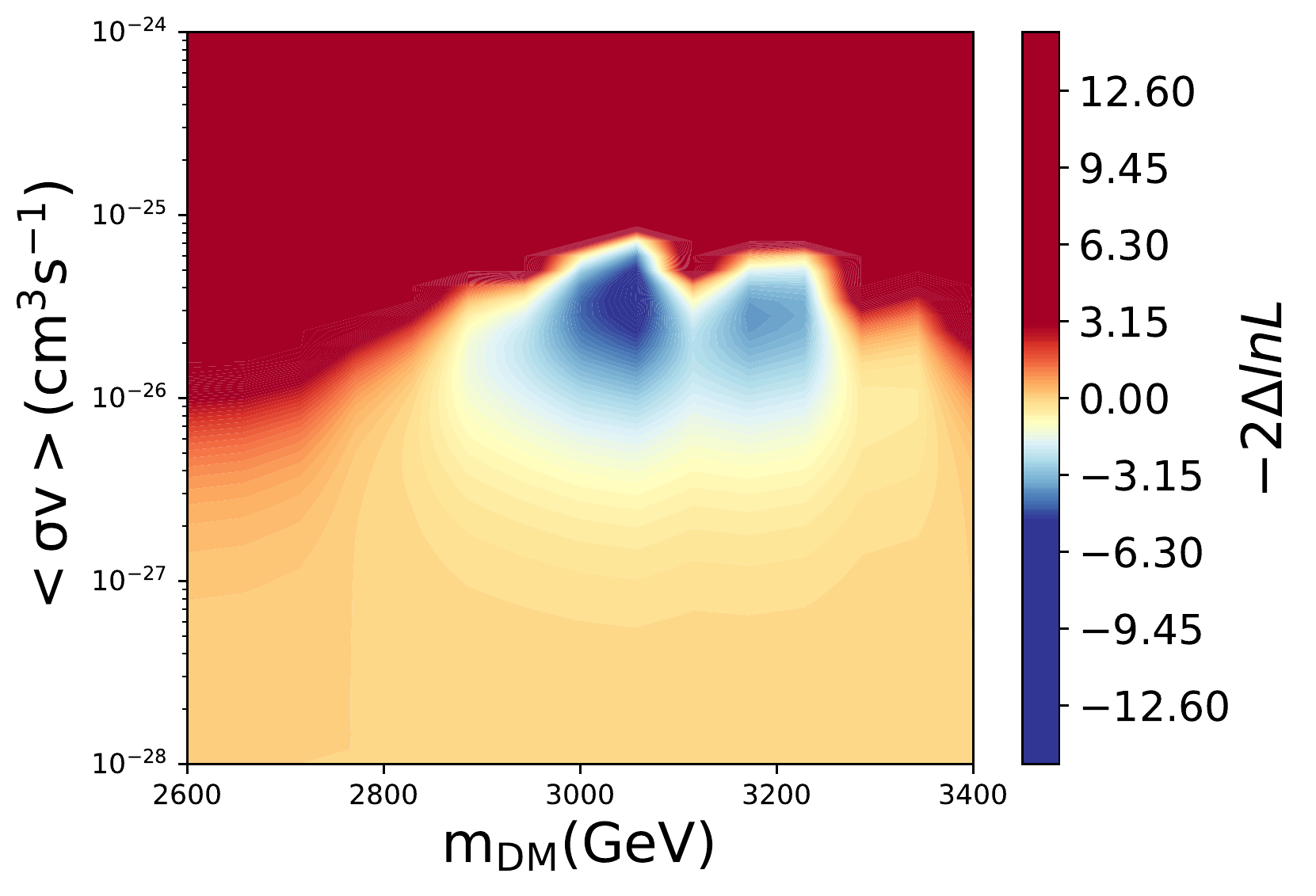}}
\caption{Log-likelihood maps on the ($\mdm,\sv$) plane in the DM subhalo
scenario, for a mass ratio of $m_{\phi}/\mdm=0.5$ (left) and $0.995$ (right).
The distance of the subhalo is $0.1~{\rm kpc}$, and its mass is
$1.9\times10^{7}$ M$_\odot$.}
\label{likelihoodlocalmap}
\end{figure}

\begin{figure}[htbp]
\centering
\includegraphics[width=0.6\columnwidth]{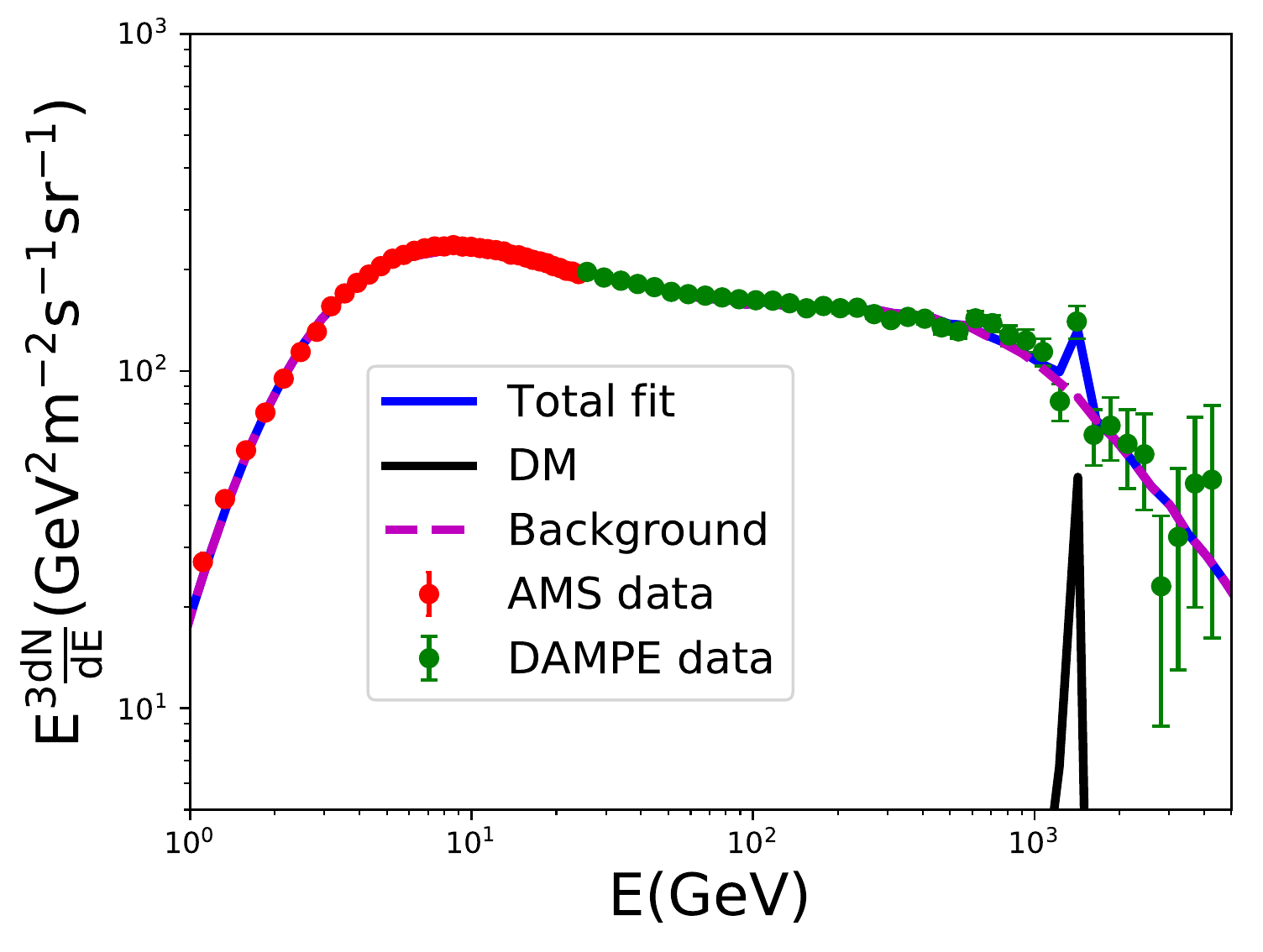}
\caption{Same as Fig.~\ref{fit} but for a local DM subhalo scenario.
Model parameters are $m_{\phi}/\mdm=0.995$, $\mdm=3~{\rm TeV}$,
$\sv=3.4\times10^{-26}~{\rm cm^3~s^{-1}}$, and the distance and mass of
the subhalo are $0.1~{\rm kpc}$ and $1.9\times10^{7}$ M$_\odot$
\cite{Yuan2017}.}
\label{localfit}
\end{figure}

Fig.~\ref{likelihoodlocalmap} shows the log-likelihood maps for such a
scenario, assuming a subhalo mass of $1.9\times10^{7}$ M$_\odot$ and a
distance of $0.1$ kpc~\cite{Yuan2017}. The left panel is for
$m_{\phi}/\mdm=0.5$, and the right panel is for $m_{\phi}/\mdm=0.995$.
We find that for a relatively small mass ratio between the intermediate
particle and the DM, the addition of the DM component always makes the
fit worse. If the mass ratio is close to 1, the fit to the data can be
improved. The best-fit result is presented in Fig.~\ref{localfit} and
the corresponding DM model parameters are $m_{\phi}/\mdm=0.995$,
$\mdm=3$~TeV, and $\sv=3.4\times10^{-26}$~cm$^3$~s$^{-1}$.

\section{Conclusion}\label{sec::con}

In this work, we have investigated a specific model in which DM particles
initially annihilate into a pair of intermediate particles which then
decay into electrons and positron. The resulting electron/positron spectrum
has a box-like shape (before propagation), which may be observed in cosmic
ray electron and positron data. Using the AMS-02 and DAMPE data we have
searched for such a possible signal.

We have considered two types of background models, one is a phenomenological
model based on the assumption that the majority of the shallow spectrum
comes from astrophysical sources, and the other is a physically motivated
model. For the physical background model, we find that the data tend to
``favor" the DM particles with $\mdm\sim(100-500)$ GeV and $\sim(2-3)$ TeV.
However, we found that such results are simply due to the degeneracy
between the DM component and the pulsar component, and hence do not really
point toward the existence of DM particles.
Therefore the $95\%$ upper limits of the DM annihilation cross section have
been derived. The constraints range from $\sim 10^{-26}~{\rm cm^3~s^{-1}}$ to
$\sim 10^{-23}~{\rm cm^3~s^{-1}}$ for DM mass from 50 GeV to 10 TeV.

Within the scenario of a local DM subhalo, we find that a high mass
ratio between $m_{\phi}$ and $\mdm$ (for which the resulting $e^+e^-$
spectrum is line like) can improve the fit to the tentative peak at
1.4 TeV. DAMPE was designed to work for 3 years~\cite{dampe-detector}.
Given its current perfect status, it is expected to work for at least
5 years. Then the statistics of electrons and positrons can be recorded
by DAMPE will increase by several times. More accurate spectrum may
shed further light on the indirect detection of DM annihilation.

\acknowledgments
This work is supported in part by the National Key Research and Development
Program of China (No. 2016YFA0400200), National Natural Science of China
(Nos. 11525313, 11722328, 11773075, U1738206), the Youth Innovation
Promotion Association of Chinese Academy of Sciences (No. 2016288),
the Natural Science Foundation of Jiangsu Province (No. BK20151608),
and the 100 Talents program of Chinese Academy of Sciences.

\end{document}